\documentclass[11pt]{article}
\usepackage{jheppub}
\usepackage{dsfont}
\usepackage{bbm}
\usepackage{slashed}
\numberwithin{equation}{section}

\newcommand{\vy}{\vec{y}}

\newcommand{\be}{\begin{equation}}
\newcommand{\ee}{\end{equation}}
\newcommand{\bea}{\begin{eqnarray}}
\newcommand{\eea}{\end{eqnarray}}

\newcommand{\vx}{\vec{x}}

\newcommand{\vp}{\vec{p}}

\newcommand{\vq}{\vec{q}}

\newcommand{\vk}{\vec{k}}

\title{Entanglement entropy in particle decay.}

\author{Louis Lello,}

\author{Daniel Boyanovsky,}

\affiliation{Department of Physics and
Astronomy, University of Pittsburgh, Pittsburgh, PA 15260}

\author{Richard Holman}

\affiliation{Department of Physics,
Carnegie Mellon University,
Pittsburgh PA 15213}

\emailAdd{lal81@pitt.edu}

\emailAdd{boyan@pitt.edu}

\emailAdd{rh4a@andrew.cmu.edu}

\abstract{The decay of a parent particle into two or more daughter particles results in an entangled quantum state  as a consequence of conservation laws in the decay process. Recent experiments at Belle and BaBar take advantage of quantum entanglement and the correlations in the time evolution of the product particles to study CP and T violations.
 If one (or more) of the product particles are not observed, their degrees of freedom are traced out of the pure state density matrix resulting from the decay, leading to a mixed state density matrix and an entanglement entropy. This entropy is a measure of the loss of information present in the original quantum correlations of the entangled state.
 We use the Wigner-Weisskopf method to construct an approximation to this state that evolves in time in a {\em manifestly unitary} way. We then obtain the entanglement entropy from the reduced density matrix of one of the daughter particles   obtained by tracing out the unobserved states, and follow its time evolution. We find that it grows over a time scale determined by the lifetime of the parent particle to a maximum, which when the width of the parent particle is narrow, describes the phase space distribution of maximally entangled Bell-like states. The method is generalized to the case in which the parent particle is described by a wave packet localized in space. Possible experimental avenues to measure the entanglement entropy in the decay of mesons at rest are discussed.}
\keywords{}
\arxivnumber{}

\begin{document}

\maketitle

\section{Introduction}

Once described as the source of ``spooky action at a distance'' by Einstein, Podolsky and Rosen (EPR)\cite{epr}, quantum entanglement has now come to be viewed as a resource to be exploited in a number of venues. It serves as the workhorse for quantum computations\cite{expt1,photatom,dutt,bookqiqc,preskill,horodecki,esentan}, and is at the heart of current efforts on quantum information. In condensed matter and quantum optics, the spontaneous decay of excited atomic states leads to quantum entangled states of photons and atoms or spin-qubits, which can then be implemented as platforms for quantum computing\cite{photatom,dutt} by transmitting the information stored in the quantum correlations of the entangled states. Furthermore, current experiments in high energy physics are beginning to exploit the quantum correlations of pairs of particles produced in meson decay to study various aspects of CP violation and time reversal invariance.

The Belle collaboration\cite{belle} has reported on remarkably precise measurements of (EPR) entanglement and correlations in  $\Upsilon(4S)\rightarrow B^{0}\overline{B}^{0}$ decays via the analysis of the time dependence of the flavor asymmetry. The BaBar collaboration\cite{trevbo,babar} has reported on the first direct measurement of time reversal violation in the  $B^{0}\overline{B}^{0}$ system from $\Upsilon(4S)$ decay at rest by studying the correlations between the members of the entangled $B^{0}\overline{B}^{0}$ pairs combining ``flavor tagging'' with ``CP tagging''. The possibility has also been advanced\cite{tviolent,soni,alok} of using entanglement correlations  to establish bounds on the $B_s-\overline{B}_s$ width difference and CP violating phases. Entanglement between the charged lepton and its associated neutrino in the decay of pseudoscalar mesons has been recently argued to play an important role in the coherence (and decoherence) aspects of neutrino oscillations\cite{nu1,patkos,smirnov,nuestro} with potentially important corrections in short baseline oscillation experiments\cite{nu2}.

In the experiments that explore CP and or T violation from the time evolution of entangled states of $B^{0}\overline{B}^{0}$ pairs the main role played by entanglement is that the information contained in the quantum correlations of the entangled states is used to carry out measurement or ``tagging'' on one or both of the members. This article will explore a complementary aspect of entanglement and correlations, namely what happens when one or more members of an entangled state cannot (or will not) be measured, so that some of the information contained in the original quantum correlations of the entangled state is lost.

Given a pure quantum system consisting of entangled subsystems, it may not be possible to measure the separate state of all of the subsystems (or while possible, we may opt {\em not} to measure them). We can then construct a {\em reduced} density matrix for the subsystem(s) we do measure by tracing over the allowed states of the unobserved subsystem(s). This then leads directly to the concept of entanglement entropy: this is the von Neumann entropy of the reduced density matrix. It reflects the loss of information that was originally present in the entangled state from the quantum correlations. This entanglement entropy has been the focus of several studies in statistical and quantum field theories  where subsystems are spatially correlated across boundaries by tracing over the degrees of freedom of one part of the system\cite{cardy,plenio,wilczek,hertzberg,srednicki,huerta} and have been extended to the case  including black holes\cite{solo,ted}, particle production in time dependent backgrounds\cite{hu} and cosmological space times\cite{maldacena}. Momentum space entanglement and renormalization has been recently studied in ref.\cite{bala}.

While entanglement entropy has been mostly studied within the context of a quantum system subdivided by space-like regions (see references above), in this article we study the time evolution of the entanglement entropy in the ubiquitous case of particle decay. The construction of the relevant states in this case relies on the Wigner-Weisskopf theory of spontaneous emission\cite{ww,book1}, which provides a \emph{non-perturbative} method for obtaining the quantum state arising from spontaneous decay. The knowledge of the full quantum state  can then be used to obtain the entanglement entropy contained in, for example, the photon-spin qubit correlations generated from the dynamics of  spontaneous decay in solid state systems\cite{mio}.

Recently this theory was generalized to relativistic quantum field theory to yield insight into the quantum states from particle decay in cosmology\cite{desitter}, and to describe potential decoherence effects in neutrino oscillations in short baseline experiments\cite{nuestro,nu2}.

\textbf{Motivation and goals:}

Motivated by   experiments at Belle and Babar that take advantage of the quantum correlations in entangled $B^{0}\overline{B}^{0}$ pairs produced  from the decay at rest of the $\Upsilon(4S)$ resonance and by proposals to study CP violating phases by exploiting correlations in entangled $B_s\overline{B}_s$ pairs, we focus on a \emph{complementary} aspect of entanglement, namely quantifying the loss of information if one of the members of the entangled pair is not measured. Tracing over the degrees of freedom of one of the members of an entangled pairs leads to a \emph{reduced} density matrix, from which we obtain the entanglement or Von-Neumann entropy which \emph{could}  potentially be a useful tool to infer correlations even when particles in the final state are unobserved.

Here we extend and generalize the Wigner-Weisskopf method discussed in\cite{nu2,desitter,mio} to describe particle decay in quantum field theory and apply it to the simple case of a bosonic parent particle decaying into two bosonic daughter particles to highlight the main consequences, although we argue that the results are general.  We address the important aspect of unitarity and obtain the entanglement entropy by tracing over the degrees of freedom associated with an unobserved daughter particle. We also show how unitary time evolution yields an entanglement entropy that grows over the lifetime of the parent particle and saturates to the entropy of maximally entangled states. Furthermore we extend the treatment to \emph{wave packets} and compare with the case wherein the quantum states are described as plane waves,  assess the corrections  and suggest a potential way to experimentally measure the entanglement entropy from the decay of mesons at rest.

\section{The Wigner-Weisskopf Method}\label{ww}

Consider a system described by a total Hamiltonian that can be decomposed as $H = H_0 + H_i$, where $H_0$ is the free Hamiltonian and $H_i$ is the interaction part. As usual, the time evolution of a state in the interaction picture is given by
\be
i \frac{d}{dt} | \Psi(t) \rangle  =  {H}_I(t) | \Psi(t) \rangle  \label{timevol}
\ee where
\be  {H}_I(t) = e^{i {H}_0 t}  {H}_i e^{-i {H}_0 t} \label{Hint}\ee is the interaction Hamiltonian in the interaction picture. The formal solution of (\ref{timevol}) is given by

\be
| \Psi(t) \rangle  =  {U}(t,t_0) | \Psi(t_0) \rangle ~~;~~ {U}(t,t_0) = T(e^{-i\int_{t_0}^{t}  {H}_I(t') dt'}) \label{psievolt}
\ee

Expanding the state $| \Psi( t) \rangle  $ in the basis of eigenstates of $H_0$  we have
\be
|\Psi(t) \rangle  = \sum_{n} C_n(t) | n \rangle \,,\label{qstate}
\ee
where $ {H}_0 | n \rangle = E_n |n \rangle$. From eq.(\ref{timevol}), upon expanding in basis states $|n\rangle$, it follows that

\be
\frac{d C_{n}(t)}{dt}  = -i \sum_{m} \langle n | H_{I}(t) | m \rangle C_{m}(t)\,. \label{evolcs}
\ee

This is an infinite set of differential equations that can be solved hierarchically leading in general to integro-differential equations at any given order in the perturbative expansion. Progress can be made by considering the evolution of the states that are coupled at a given order in perturbation theory and solving the coupled equations for these, thereby truncating the hierarchy at a given order in the perturbative expansion.

For the case of interest here, consider an initial value problem in which the system is prepared at an initial time $t=0$ in a  state $|A\rangle$, so that $C_A(0)=1,\ C_{n\neq A} =0$ and that the Hamiltonian $H_I$ couples these states to a set of states $|\kappa\rangle$. Then, we can close the hierarchy at second order in the interaction by keeping only the coupling of the states $|A\rangle  \leftrightarrow |\kappa\rangle$, i.e.
 \be
\frac{d}{dt}\, C_{A}(t) = -i \sum_{\kappa} \langle A | H_{I}(t) | \kappa \rangle \, C_{\kappa}(t)\label{Meq}
\ee

\be
\frac{d}{dt}\, C_{\kappa}(t) = -i \langle \kappa | H_{I}(t) |A \rangle \, C_{A}(t)\,. \label{kapaeqn}
\ee

Using the initial conditions we obtain

\be
C_{\kappa}(t) = -i \int_{0}^{t} dt' \langle \kappa | H_{I}(t') | A \rangle C_{A}(t')\label{ckaps}
\,, \ee which when inserted into (\ref{Meq}) leads to

\be
\frac{d}{dt} C_{A}(t) = - \int_{0}^{t} dt' \sum_{\kappa} \langle A | H_{I}(t) | \kappa \rangle \langle \kappa | H_{I}(t') | A \rangle C_{A}(t') = - \int_{0}^{t} dt' \Sigma_A(t-t') C_A(t')\,,\label{meq2}
\ee
where the  second order self energy has been introduced

\be
\Sigma_{A}(t-t') \equiv \sum_{\kappa} \langle A | H_{I}(t) | \kappa \rangle \langle \kappa | H_{I}(t') | A \rangle =  \sum_{\kappa} \Big| \langle A | H_{I}(0) | \kappa \rangle \Big|^2~ e^{i(E_A-E_\kappa)(t-t')}  \,.\label{sigmaA} \ee

Higher order corrections can be included by enlarging the hierarchy, i.e. by considering the equations that couple the states $\kappa$ to other states $\kappa'$ via the Hamiltonian. The coefficients for the states $\kappa'$ can be obtained by integration and can be inserted back in the equations for the coefficients $C_\kappa$ (which already include their coupling to $|A\rangle$). Then formally integrating the equation and inserting the results back into the equation for $|A\rangle$ generates higher order corrections to the self-energy $\Sigma_A$. Finally, solving for the time evolution of $C_A$ allows us to obtain the time evolution of the other coefficients.

As seen by the procedure described above, the Wigner-Weisskopf approach can be used to construct an approximate version of the quantum state in the presence of interactions. However, what is not altogether obvious is that the truncation of states used to construct the state gives rise to a state whose time evolution is {\em unitary}. This will be extremely important in the sequel since we will want to follow that time evolution of the entanglement entropy this state would provide after tracing out an unobserved subsystem as discussed in the introduction. We will need to be sure that there are no spurious effects in this evolution due to an   approximation to the state.

The statement of unitary is one of conservation of probability. From the evolution equation (\ref{evolcs}) and its complex conjugate it follows that
\be \frac{d}{dt}\sum_{n}|C_n(t)|^2 = -i \sum_{m,n}\Bigg[C_m(t)C^*_n(t)\langle n|H_I(t)|m\rangle-C_n(t)C^*_m(t)\langle m|H_I(t)|n\rangle \Bigg]  =0 \,, \label{unitacs}\ee as can be seen by relabeling $m\leftrightarrow n$ in the second term.  Therefore $\sum_{n}|C_n(t)|^2 = \mathrm{constant}$. Now this is an \emph{exact} result; the question is whether and how is it fulfilled in the Wigner-Weisskopf approximation obtained by truncating the hierarchy to the set of equations (\ref{Meq},\ref{kapaeqn}).

Using eqs.(\ref{ckaps}, \ref{sigmaA}) consider
\be \sum_{\kappa}|C_\kappa(t)|^2 = \int_0^{t}dt_1 C^*_A(t_1)\int_0^{t}dt_2\Sigma_A(t_1,t_2) C_A(t_2). \label{sumkapa}\ee Inserting $1=\Theta(t_1-t_2)+\Theta(t_2-t_1)$ in the time integrals it follows that
\bea \sum_{\kappa}|C_\kappa(t)|^2 & = &  \int_0^{t}dt_1 C^*_A(t_1)\int_0^{t_1}dt_2\Sigma_A(t_1,t_2) C_A(t_2)  \nonumber \\ & + & \int_0^{t}dt_2 C_A(t_2)\int_0^{t_2}dt_1\Sigma_A(t_1,t_2) C^*_A(t_1)\label{shuffle} \eea so that using $\Sigma_A(t_1,t_2) = \Sigma^*_A(t_2,t_1)$ relabelling $t_1 \leftrightarrow t_2$ in the second line of (\ref{shuffle}) and using (\ref{meq2}) we find
\bea  &&\sum_{\kappa}|C_\kappa(t)|^2   =   - \int_0^{t}dt_1 \Big[ C^*_A(t_1)\frac{d}{d t_1}C_A(t_1) + C_A(t_1) \frac{d}{d t_1}C^*_A(t_1)\Big] = - \int_0^{t}dt_1 \frac{d}{dt_1} |C_A(t)|^2\nonumber\\
&& = 1-|C_A(t)|^2 \label{unita}\eea where we have used the initial condition $C_A(0)=1$. This is the statement of unitary time evolution, namely
\be  |C_A(t)|^2 + \sum_{\kappa}|C_\kappa(t)|^2 = |C_A(0)|^2 \label{unitime}\ee

This is an important result. As we will see below, standard perturbative calculations of this state would {\em not} yield a state that evolved unitarily. The Wigner-Weisskopf state involves a non-perturbative dressing up of the state and despite its approximate nature, this dressing up captures the physics to a sufficient extent to guarantee unitary time evolution.

\section{Reduced density matrix and entanglement entropy.}\label{sec:enta}

Now we are ready to turn to the problem we really want to consider: the state that appears after the decay of a parent particle $\Phi$ into two daughters $\chi ,\psi $. For simplicity and to highlight the main concepts we treat   all fields as bosonic massive fields with masses $m_\Phi$   and $m_\chi, m_\psi$ respectively. We consider a typical interaction vertex described by the interaction Hamiltonian
\be H_I = g \int d^3 x \, \Phi(\vec{x})\,\chi (\vec{x})\,\psi (\vec{x})\,. \label{Hi}\ee

We quantize the fields in a volume $V$, so that in the interaction picture they can be written as:
\be \varphi(\vec{x},t) =  \sum_{\vk}\frac{1}{\sqrt{2E_k V}}\Big[a_{\vk} \,e^{-iE_k t} \,e^{i\vk\cdot\vec{x}}+ a^\dagger_{\vk} \,e^{iE_k t} \,e^{-i\vk\cdot\vec{x}}\Big]~~;~~\varphi = \Phi, \chi , \psi  \,. \label{expansion} \ee

\begin{figure}[h!]
\begin{center}
\includegraphics[keepaspectratio=true,width=3in,height=3in]{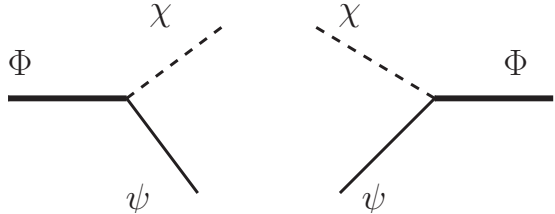}
\caption{Transitions $| \Phi \rangle \leftrightarrow |\chi   \rangle | \psi    \rangle$ up to order $g^2$ that determine $\Sigma_\Phi$.}
\label{fig:wwpinumu}
\end{center}
\end{figure}

For the case of interest here, namely the decay process $\Phi \rightarrow \chi  \,\psi  $ we consider that  the initial state is a single particle $\Phi$ at rest, therefore the initial  condition is  $C_{\Phi}(\vk=0; t=0) = 1~,~C_{\kappa}(t=0) =0 ~\mathrm{for}~ |\kappa\rangle \neq |1^{\Phi}_{\vec{0}}\rangle$. The interaction Hamiltonian (\ref{Hi}) connects the initial state, $| 1^{\Phi}_{\vec{0}} \rangle$   to the states, $|\kappa\rangle =|\chi  _{-\vp}\rangle\ \otimes | \psi  _{\vp} \rangle $. These states in turn are coupled back to $| 1^{\Phi}_{\vec{0}} \rangle$ via $H_I$,  these processes are depicted in fig.(\ref{fig:wwpinumu}).

Thus to leading order in $g$ we find from the intermediate states shown in fig.(\ref{fig:wwpinumu})
\be \Sigma_{\Phi}(t-t')= \sum_{\vp} |\langle 1^{\Phi}_{\vec{0}} | \hat{H}_I(0)| \chi  _ {-\vp}, {\psi  }_{\vp} \rangle|^{2} e^{i(m_{\Phi} -E_{\chi  }(p)-E_\psi  (p))(t-t')}\,. \label{leadsigma}\ee

\begin{figure}[h!]
\begin{center}
\includegraphics[keepaspectratio=true,width=2in,height=2in]{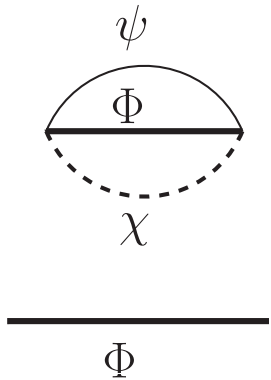}
\caption{Order $g^2$ correction to the vacuum energy. Yields an overall phase for the quantum state $|1^{\Phi}\rangle$ \cite{desitter}.  }
\label{fig:disco}
\end{center}
\end{figure}

The interaction Hamiltonian also connects a single $\Phi$-particle state to an intermediate state with three other particles and this state back to the single $\Phi$ particle state yielding a disconnected contribution to the self energy depiced in fig.(\ref{fig:disco}). This contribution is just a renormalization of  the vacuum energy and only contributes to an overall phase that multiplies the single particle $\Phi$ state and will be neglected in the following analysis. For a more detailed discussion of this contribution see ref.\cite{desitter}.

At higher order in $g$ there are higher order contributions to the $\Phi$ self energy from other multiparticle states.  However, we will only be considering states connected to $|1^{\Phi}\rangle$ to first order in perturbation theory. As discussed in detail in refs.\cite{nu2,desitter} the Wigner-Weisskopf method provides a \emph{non-perturbative resummation} of self-energies in \emph{real time} akin to the dynamical renormalization group\cite{drg}. The self-energy  eq.(\ref{leadsigma}) is recognized as the one-loop retarded self energy of the field $\Phi$ with the $|\chi  \rangle |{\psi  }\rangle$ intermediate state\cite{nu2,desitter}.

Solving eq.(\ref{meq2}) produces a solution for the time evolution of the $\Phi$ amplitude. We can use the solution for $C_\Phi(t)$ to obtain an expression for the amplitudes $C_{\kappa}(t)$ which allows for computation of the probability of occupying a particular state at any given time. We may solve eq.(\ref{meq2}) either via Laplace transform, or in the case of weak coupling, a derivative expansion which yields the same result at long times ($t\gg 1/ m_{\Phi}$). In ref.\cite{desitter} the equivalence between the two approaches is discussed in detail.  Here, we follow the latter method which is the original Wigner-Weisskopf approximation\cite{ww,book1}.

We begin by defining the quantity

\be
W_0(t,t') = \int^{t'}_{0} dt'' \Sigma_{\Phi}(t-t'')
\ee

so that

\be
\frac{d}{dt'} W_0(t,t') = \Sigma_{\Phi}(t-t')~~,~~  W_0(t,0)=0
\ee

Integrating eq.(\ref{meq2}) by parts yields

\be
\frac{d}{dt} C_{\Phi}(t) = -\int_{0}^{t} dt' \Sigma_{\Phi}(t-t')~ C_{\Phi}(t') = -W_0(t,t)~C_{\Phi}(t)+\int_{0}^{t} dt'W_0(t,t')\frac{d}{dt'}C_{\Phi}(t').
\ee

The  first is term  second order in $H_I$ whereas the second term is of  fourth order in $H_I$ and will be neglected. This approximation is equivalent to the Dyson resummation of the one-loop self energy diagrams.   Thus to leading order, eq.(\ref{meq2}) becomes

\be
\frac{d}{dt}C_{\Phi}(t) + W_0(t,t) C_{\Phi}(t) = 0 \,,\label{finmeq}
\ee  where

\be
 W_0(t,t) = \int^{t}_{0} dt' \Sigma_{\Phi}(t-t') = \int^{t}_{0} dt' \sum_{\vp} |\langle 1^{\Phi}_{\vec{0}} | \hat{H}_I(0)| \chi  _ {-\vp}, {\psi  }_{\vp} \rangle|^{2} e^{i(m_{\Phi} -E_{\chi  }(p)-E_\psi  (p))(t-t')}   \label{wott}
\ee

Inserting a convergence factor and taking the   limit $t\rightarrow \infty$ consistently with the Wigner-Weisskopf approximation, we find\footnote{The long time limit in the Wigner-Weisskopf approximation is equivalent to the Breit-Wigner approximation of a resonant propagator\cite{desitter} and holds for $t\gg 1/m_{\Phi}$.}

\be
 W_0(t,t)  = \lim_{\epsilon \to 0^{+}} i  \sum_{\kappa}\frac{ |\langle 1^{\Phi}_{\vec{0}} | \hat{H}_I(0)| \chi  _ {-\vp}, {\psi  }_{\vp} \rangle|^{2}}{m_{\Phi} -E_{\chi  }(p)-E_\psi  (p)+i\epsilon} = i \Delta E_\Phi + \frac{\Gamma }{2}
\ee where

\be \Delta E_\Phi \equiv \mathcal{P}~ \sum_{\vp}\frac{  |\langle 1^{\Phi}_{\vec{0}} | \hat{H}_I(0)| \chi  _ {-\vp}, {\psi  }_{\vp} \rangle|^{2}}{m_{\Phi} -E_{\chi  }(p)-E_\psi  (p)}\,,\label{eshift}\ee is the second order shift in the energy which will be absorbed into a renormalization of the  $\Phi$ mass and
\be \Gamma \equiv 2 \pi \sum_{\vp}  |\langle 1^{\Phi}_{\vec{0}} | \hat{H}_I(0)| \chi  _ {-\vp}, {\psi  }_{\vp} \rangle|^{2} \delta(m_{\Phi} -E_{\chi  }(p)-E_\psi  (p))\,\label{Pwidth}\ee is the decay width as per Fermi's Golden rule. Therefore  in this approximation, we arrive at

\be
C_{\Phi}(t) = e^{-i\Delta E_\Phi t}~ e^{-\frac{\Gamma }{2} t} \,, \label{Cpi}
\ee where we now consider a $\Phi$ with $\vk =0$ (decay at rest) and find
\be \mathcal{M}_{\Phi}(p) = \langle 1^\Phi_{\vec{0}} | \hat{H}_I(0)| \chi  _{\vp}\,\psi  _{-\vp}\rangle = \frac{g}{\sqrt{8Vm_{\Phi}E_{\chi  }(p)E_{\psi  }(p)}} \label{mtxele}\ee leading to
\be \Gamma= 2 \pi\sum_{p} |\mathcal{M}_\Phi(p)|^2  \delta\big(m_\Phi - E_{\chi  }(p)-E_{\psi  }(p)\big)=\frac{g^2\,p^*}{8\pi m^2_{\Phi}}\label{gamapi}\ee where
\be p^* = \frac{1}{2m_{\Phi}}\Bigg[ m^4_\Phi +m^4_\chi  +m^4_\psi   - 2 m^2_\Phi  m^2_\chi   - 2 m^2_\Phi m^2_\psi  - 2 m^2_\chi   m^2_\psi  \Bigg]^{1/2}\,. \label{pstar}\ee
Inserting (\ref{Cpi}) into (\ref{ckaps}) we find the quantum state
\be |\Psi(t)\rangle  = e^{-i\Delta E_\Phi t}~ e^{-\frac{\Gamma }{2} t}\,|1^{\Phi}_{\vec{0}};0_\chi  ;0_\psi  \rangle + \sum_{\vp} \mathcal{C}_{\chi \psi}(p\,;t)\,|\chi  _{\vp}\rangle\,|\psi  _{-\vp}\rangle |0_\Phi\rangle \label{psioftfin}\ee where\
\be  \mathcal{C}_{\chi \psi}(p\,;t) =\mathcal{M}_{\Phi}(p)~ \frac{\Big[1-e^{-i(m_{\Phi,R}-E_\chi  (p)-E_\psi  (p)-i\Gamma/2)\,t} \Big]}{\Big(E_\chi  (p)+E_\psi  (p)-m_{\Phi,R}+i\Gamma/2\Big)}\label{coef} \ee and $m_{\Phi,R} = m_{\Phi}+\Delta E_\Phi$ is the renormalized mass of the $\Phi$ particle. In what follows we drop the subscript $R$ and always refer to the renormalized mass.

At this stage we can make contact with the momentum entanglement discussion in ref.\cite{bala} and at the same time exhibit the true non-perturbative nature of the results above by considering the state obtained using naive perturbation theory.

Taking the initial state at $t=0$ to be  $|1^{\Phi}_{\vec{0}}\rangle$, then to leading order in $g$,  the time evolved state is given by
\be |\Psi(t)\rangle = \Big[1 -i \int^t_0 e^{iH_0t'} H_I e^{-iH_0t'} dt'+\cdots \Big]|1^{\Phi}_{\vec{0}}\rangle \, .\label{pert}\ee Introducing a resolution of the identity $1=\sum_{\kappa} |\kappa\rangle \langle \kappa|$ we find
\be |\Psi(t)\rangle = |1^{\Phi}_{\vec{0}}\rangle + \sum_{\vp} \Bigg[\mathcal{M}_{\Phi}(p)~ \Bigg(\frac{ 1-e^{-i(m_{\Phi}-E_\chi  (p)-E_\psi  (p) )\,t}  }{E_\chi  (p)+E_\psi  (p)-m_{\Phi} }\Bigg) \Bigg] |\chi  _{-\vp}\rangle\,|\psi  _{\vp}\rangle |0_\Phi\rangle +\cdots \,.\label{psipert}\ee  In this expression we have neglected a disconnected three particle intermediate state in which the initial  $|1^{\Phi}_{\vec{0}}\rangle$ remains and the interaction creates an intermediate state with three particles. This contribution is truly perturbative, does not contain resonant denominators as (\ref{psipert}) and corresponds to the disconnected diagram in fig.(\ref{fig:disco}).

The probability of finding the daughter states is given by the familiar result (Fermi's Golden rule)
\be \mathcal{P}(t) = \sum_{\vp} \big|\mathcal{M}_{\Phi}(p)\big|^2\, \Bigg[\frac{\sin\big((m_{\Phi}-E_\chi  (p)-E_\psi  (p))\,t/2\big)}{(m_{\Phi}-E_\chi  (p)-E_\psi  (p))}\Bigg]^2 = \Gamma t \label{probaFG}\ee where $\Gamma$ is given by (\ref{gamapi}). This result is obviously only valid for $t\ll 1/\Gamma$. It is now clear that the generalized Wigner-Weisskopf method that yields the state (\ref{psioftfin}) with the coefficients given by (\ref{coef}) is truly non-perturbative.

 The momentum entanglement between the daughter particles is akin to that discussed perturbatively in ref.\cite{bala} with some  important differences. In ref.\cite{bala} momentum entanglement was studied for the vacuum wave function. The corresponding contributions are truly perturbative and do not feature the resonant denominators that lead to secular growth in time and are  similar to the contributions that we neglect and are described by fig.(\ref{fig:disco}).

The quantum state (\ref{psioftfin}) describes an \emph{entangled} state of the parent and daughter particles. The full (pure state) density matrix is given by
\be \rho(t) = |\Psi(t)\rangle\langle \Psi(t)| \label{rho}\ee and its trace is given by
\be \mathrm{Tr}\rho(t) = e^{-\Gamma t}+ V \int \frac{d^3p}{(2\pi)^3} |\mathcal{C}_{\chi \psi}(p\;;t)|^2 \label{tracerho}\ee

The momentum integral in (\ref{tracerho}) can be computed by changing variables to $\mathcal{E} = E_\chi  (p)+E_\psi  (p)$. In the narrow width limit $\Gamma \ll m_{\Phi}, m_\chi  +m_\psi  $, the integrand is sharply peaked at $\mathcal{E} = m_{\Phi}$ so that the lower limit can be consistently taken to $-\infty$  thus allowing the integral can be computed  by contour integration. We find
\be V \int \frac{d^3p}{(2\pi)^3} |\mathcal{C}_{\chi \psi}(p\,;t)|^2 = 1-e^{-\Gamma t}\,, \label{intC}\ee confirming that
\be \mathrm{Tr}\rho(t) =1\,, \label{unitrace}\ee consistent with unitary time evolution and the unitarity relation (\ref{unitime}). Furthermore   the average number of $\psi  $ (or $\chi  $) particles is given by
\be n^{\psi  }(p\,;t) = \langle \Psi(t)|a^\dagger_{\psi  }(p) a_\psi  (p) |\Psi(t)\rangle \equiv (2\pi)^3 \frac{dN^{\psi  }(t)}{d^3xd^3p} = |\mathcal{C}_{\chi \psi}(p\,;t)|^2\,, \label{nofnu}\ee thus the \emph{total} number of $\psi  $ (or $\chi  $) particles is
\be N^\psi  (t) = V\int \frac{d^3p}{(2\pi)^3}\, n^{\psi  }(p\,;t)=  1-e^{-\Gamma t}\,. \label{Nnutot}\ee Tracing out one of the daughter particles, for example $\chi$ if it is unobservable, leads to a mixed state \emph{reduced density matrix}
\be \rho_\psi  (t)  = \mathrm{Tr}_{\chi  } \rho(t) = e^{-\Gamma t}\, |1^{\Phi}_{\vec{0}}\rangle \langle 1^{\Phi}_{\vec{0}}| + \sum_{\vp}  |\mathcal{C}_{\chi \psi}(p\,;t)|^2\,  |\psi  _{\vec{p}}\rangle \langle\psi  _{\vec{p}}|\,. \label{rhonu}\ee The Von-Neumann entanglement entropy is therefore given by
\be S(t) =  -n^\Phi(0,t)\,\ln[n^\Phi(0,t)] - \sum_{\vp} n^{\psi  }(p\,;t) \ln \big[n^{\psi  }(p\,;t)\big] \,. \label{entropy}\ee where $n^{\psi  }(p)$ is given by (\ref{nofnu}) and  $n^\Phi(0,t)= e^{-\Gamma t} $.   Because in the narrow width limit $|\mathcal{C}_{\chi \psi}(p\,;t)|^2$ is a sharply peaked distribution, under integration with functions that vary smoothly near $p \simeq p^*$ it can be replaced by
\be n^{\psi  }(p\,;t) = |\mathcal{C}_{\chi \psi}(p\,;t)|^2 \simeq  \frac{2\pi^2 m_{\Phi}\,\big[1-e^{-\Gamma t}\big]}{V p^* E_\chi  (p) E_\psi  (p)} ~  \frac{1}{2\pi} \,  \frac{\Gamma }{(E_\chi  (p)+E_\psi  (p)-m_{\Phi})^2+(\Gamma/2)^2} \,. \label{Capp}\ee Using this approximation we find
\be S(t) =  \Gamma t e^{-\Gamma t}- \big[1-e^{-\Gamma t}\big]\ln\big[1-e^{-\Gamma t}\big] - \big[1-e^{-\Gamma t}\big]\ln\big[n^{\psi  }(p^*\,;\infty)\big]\label{entrofin}\ee
where
\be n^{\psi  }(p^*\,;\infty) = \frac{4\pi \, m_{\Phi}}{V p^* E_\chi  (p^*) E_\psi  (p^*)\Gamma}\,. \label{ninfty}\ee
and the asymptotic entanglement entropy is
\be S(\infty) =  -\ln\Bigg[\frac{4\pi \, m_{\Phi}}{V p^* E_\chi  (p^*) E_\psi  (p^*)\Gamma}\Bigg]\,.\label{entroasy}\ee

This result has the following interpretation. In the asymptotic limit $t \gg 1/\Gamma$, the entanglement (von Neumann) entropy approaches (minus) the logarithm of the available states. The decay of the parent particle produces entangled pairs in which each member features a very narrow distribution centered at $p^*$ of width $\sim \Gamma$ and height $\sim 1/\Gamma$. The total area in momentum space yields $1/V$ since there is only one particle (of either type) produced in the volume $V$. Within the range of momenta centered at $p^*$ and of width $\Gamma$ all of the available single particle states have equal probability $\propto 1/V$, therefore these states are \emph{maximally entangled} as Bell states.  This observation becomes clearer recognizing that a typical quantum state  that contributes to the sum in (\ref{psioftfin})  is of the form
\be \mathcal{C}_{\chi \psi}(p^*,t) \Big[|\chi  (\vec{p^*})\rangle \,|\psi  (-\vec{p^*})\rangle + |\chi  (-\vec{p^*})\rangle \, |\psi  (\vec{p^*})\rangle \Big] ~~;~~ \vec{p^*} = p^*\,\vec{n}\,, \label{Bell}\ee where $\vec{n}$ is the direction of emission of either member of the pair. The quantum states with momenta $p^*-\Gamma/2 \leq p \leq p^*+\Gamma/2$ are represented with  nearly the same probability $|\mathcal{C}_{\chi \psi}(p^*,t)|^2\propto 1/V\Gamma$ in the sum. These states are Bell-type states and are maximally entangled, in fact these are similar, up to an overall normalization,  to the entangled $B^0 \overline{B^0}$ states resulting from the decay of the  $\Upsilon(4S)$ resonance\cite{belle,trevbo,babar}, but with the opposite  relative sign because of charge conjugation. If the decaying particle has a short lifetime corresponding to a broad resonance, the emitted pairs will feature a distribution of momenta with probabilities $|\mathcal{C}_{\chi \psi}(p,t)|$ determined by the Lorentzian profile of the resonance.

As discussed above $ n^{\psi  }(p^*\,;\infty)$ (see eqn. (\ref{nofnu}))  is the asymptotic phase space density of the produced particle (either $\chi  $ or $\psi  $). The entanglement entropy vanishes at the initial time since the density matrix at $t=0$ is a pure state and grows to its asymptotic value on a time scale $1/\Gamma$.

\section{Wave packets}\label{wavepack}

The treatment above described parent and daughter particles in terms of single particle plane waves, however in typical experiments the parent particle is produced as a wave packet with some localization length scale determined by the experimental setup. In this section we extend the treatment to a wave packet description. We use the discrete momentum representation in a quantization volume $V$.

Consider a particle of species $\alpha=\Phi,\chi,\psi$, Fock states describing single particle plane wave states of momentum $\vk$,   $|1^{\alpha}_{\vk}\rangle$, are normalized such that
\be \langle 1^{\alpha}_{\vk}|1^{\alpha}_{\vk'}\rangle = \delta_{\vk,\vk'} \label{normalization}\,.\ee
 Localized single particle states are constructed as linear superpositions
\be |\alpha;\vk_0,\vx_0\rangle = \sum_{\vk}C_\alpha(\vk;\vk_0;\vx_0)\,|1^{\alpha}_{\vk}\rangle \label{localsp}\ee where $C_\alpha(\vk;\vk_0;\vx_0)$ is the   amplitude, normalized so that
 \be \langle \alpha;\vk_0,\vx_0 | \alpha;\vk_0,\vx_0\rangle = \sum_{\vk}|C_\alpha(\vk;\vk_0;\vx_0)|^2=1 \,.\label{normawp}\ee

 The total number of particles in the volume $V$ is
 \be  \langle \alpha;\vk_0,\vx_0|\sum_{\vk}a^{\dagger}_{\alpha,\vk} a_{\alpha,\vk}|\alpha;\vk_0,\vx_0\rangle = 1 \,.\label{numerouno}\ee

 For a monochromatic plane wave $C_\alpha(\vk;\vk_0;\vx_0)= \delta_{\vk,\vk_0}$. The spatial wave function corresponding to the wave packet is given by
 \be \Xi(\vec{x}) = \frac{1}{\sqrt{V}} \sum_{\vk}C_\alpha(\vk;\vk_0;\vx_0)\,e^{-i\vk\cdot\vx} \,, \label{wafu}\ee the normalization (\ref{normawp}) implies
 \be \int d^3 x |\Xi(\vx)|^2 =1 \,.\label{normawafu}\ee For a monochromatic plane wave it follows that $\Xi(\vx)$ is a volume normalized plane wave.

 The average momentum of the wave packet state is given by
 \be \langle \alpha;\vk_0,\vx_0 |\sum_{\vk} \vk \,a^{\dagger}_{\alpha,\vk} a_{\alpha,\vk} |\alpha;\vk_0,\vx_0\rangle = \sum_{\vk}\vk |C_\alpha(\vk;\vk_0;\vx_0)|^2 \label{kave} \ee where $a^{\dagger}_{\alpha,\vk}; a_{\alpha,\vk}$ are the creation and annihilation operators for the species $\alpha$. Assuming that the distribution $C_\alpha(\vk;\vk_0;\vx_0)$ is isotropic in the rest frame of the wave-packet and that the average momentum is $\vk_0$, it follows that
 \be C_\alpha(\vk;\vk_0;\vx_0) = C_\alpha(\vk-\vk_0;\vx_0)\,. \label{propk}\ee

As a specific example we consider Gaussian wave packets,
\be   C_\alpha(\vk-\vk_0;\vx_0) = \Bigg[\frac{8\,\pi^\frac{3}{2}}{\sigma^3\,V} \Bigg]^\frac{1}{2}~e^{-\frac{(\vk-\vk_0)^2}{2\sigma^2}}~e^{i(\vk-\vk_0)\cdot\vx_0} \label{gaussianwf}\,,\ee where $\sigma$ is the localization in momentum space, the spatial wave function is
\be \Xi(\vx) = \Bigg[\frac{\sigma}{\sqrt{\pi}}\Bigg]^{3/2}\, e^{-i\vec{k}_0\cdot \vec{x}}\,e^{-\frac{\sigma^2}{2}(\vec{x}-\vec{x}_0)^2} \,.\label{psigau}\ee The spatial wave function is localized at $\vec{x}_0$ with  localization length   $1/\sigma$ and the momentum wave function is localized at $\vec{k}_0$ which is the average momentum in the wave packet  and the momentum localization scale is $\sigma$. The plane wave limit is obtained by formally identifying $\sigma/\sqrt{\pi} \rightarrow 1/V^{1/3}~~;~~V \rightarrow \infty$.

\vspace{2mm}

\subsection{Macroscopic localization and orthogonality:}\label{sec:local}

 In terms of these wave functions   the overlap of two wave packets with different momenta localized at different spatial points is
\be \langle \alpha;\vq_0;\vy_0|\alpha;\vk_0;\vx_0\rangle =  e^{-\frac{(\vk_0-\vq_0)^2}{4\sigma^2}}~ e^{-\frac{\sigma^2}{4}(\vx_0-\vy_0)^2}~e^{-\frac{i}{2}(\vk_0-\vq_0)\cdot(\vx_0-\vy_0)} \label{overlap}\,.\ee

For a macroscopic localization length $1/\sigma$ the wavepackets are nearly orthogonal in momentum for values of the momentum of experimental relevance. For example, consider a localization length $\simeq 1\,\textrm{meter}$
\be 1/\sigma \simeq 1~\mathrm{meter} \Rightarrow \sigma \simeq 2\times 10^{-7} \,\mathrm{eV} \label{macro} \ee whereas in typical experiments $k_0, q_0 \gtrsim \mathrm{MeV}$ in particular, for the decay of mesons, if the localization length scale is of the order of the decay length the typical ratio $k_0/\sigma, q_0/\sigma \gtrsim 10^{12}$ and typical energy (momentum) resolutions are $\gg \sigma$. Therefore for all experimental intent and purpose the wave packets are orthogonal for different values of the momentum
\be \langle \alpha;\vq_0;\vx_0|\alpha;\vk_0;\vx_0\rangle \simeq \delta_{\vq_0,\vk_0}\,. \label{ortomom}\ee


From  the identity (\ref{propk}) an important property of these wave packets that will be useful below is the following identity,
\be \sum_{\vk}C_\alpha(\vk-\vk_0;\vx_0)\,|1^{\alpha}_{\vk-\vq}\rangle = |\alpha;\vk_0-\vq;\vx_0\rangle \,.\label{prop}\ee Although this property is evident with the Gaussian wave packets (\ref{gaussianwf}) it is quite general   for localized functions of $\vk-\vk_0$.

\vspace{2mm}


 \subsection{Time evolution:}\label{sec:timeevol} Consider the single particle wavepacket (\ref{localsp}) with $C_\alpha(\vk;\vk_0;\vx_0) $ given by (\ref{gaussianwf}). The time evolution of this state is given by
\be |\alpha;\vk_0;\vx_0;t\rangle = e^{-iH_0t}|\alpha;\vk_0;\vx_0\rangle = \sum_{\vk}C_\alpha(\vk-\vk_0;\vx_0)\,e^{-iE_{\alpha}(k)t} |1^{\alpha}_{\vk}\rangle \,.\label{timevolwp}\ee For a wavepacket sharply localized in momentum, we can expand
\be E_{\alpha}(k) = E_{\alpha}(k_0) + \vec{V}_g(k_0)\cdot(\vk-\vk_0) + \cdots\,, \label{enerexp}\ee where
\be \vec{V}_g(k_0) = \frac{\vec{k}_0}{E_\alpha(k_0)} \label{groupvelo} \,\ee is the group velocity, the second derivative terms in (\ref{enerexp}) give rise to transverse and longitudinal dispersion. Neglecting both transverse and longitudinal dispersion under the assumption that the packet is narrowly localized in momentum and the time scales are much shorter than the dispersion scales (see discussion below), it is straightforward to find
\be |\alpha;\vk_0;\vx_0;t\rangle = e^{-iE_\alpha(k_0)t} \,|\alpha;\vk_0;(\vx_0-\vec{V}_g(k_0)t)\rangle \,, \label{centroid}\ee namely, neglecting dispersion the center of the wave packet moves with the group velocity as expected.


\vspace{2mm}

\subsection{N-particle wavepackets:}\label{sec:Nparts} We have normalized the wavepackets to describe one particle in the volume $V$ as is evident from the result (\ref{numerouno}). However, for experimental purposes one may consider initial states with a single particle within the \emph{localization volume } $1/\sigma^3$. Because the quantum field theory is quantized in a (much larger) volume $V$, these initial states must, therefore, feature different normalization. This can be seen from the normalization of the
single particle states (\ref{normawp}) with the usual passage to the continous momentum description
\[ \sum_{\vk} \rightarrow  V \int \frac{d^3k}{(2\pi^3)} \]
which explains the volume factor in (\ref{gaussianwf}).
The condition of one single particle state per localization volume therefore requires a different normalization of the single particle wavepackets, which can be obtained by dividing up the total volume $V$ into bins of volume $1/\sigma^3$ with one single particle in each bin, leading to $N= V\sigma^3$ total particles in the volume V, with a particle density in the volume V given by $N/V=1/(1/\sigma^3)=\sigma^3$, so that the total normalization in the whole volume $V$  must be $N$.

Therefore, we can accomplish the description of the initial state of the decaying parent particle in terms of  a wavepacket of single particle states with one single particle within the localization volume by normalizing these states to   $N$ in the   volume $V$. Namely we consider the initial wavepacket states in terms of the momentum wavefunctions
\be C_{N\alpha}(\vk-\vk_0;\vx_0) = \sqrt{N}\, C_\alpha(\vk-\vk_0;\vx_0)\,, \label{Nparwp}\ee in which case the Gaussian wavepacket (\ref{gaussianwf}) becomes
\be  C_{N\alpha}(\vk-\vk_0;\vx_0) = \Bigg[\frac{8\,\pi^\frac{3}{2}\,n}{\sigma^3} \Bigg]^\frac{1}{2}~e^{-\frac{(\vk-\vk_0)^2}{2\sigma^2}}~e^{i(\vk-\vk_0)\cdot\vx_0}\,, \label{Ngaus} \ee where $n = N/V=\sigma^3$ is the particle \emph{density}. The corresponding quantum states
\be |N\alpha;\vk_0,\vx_0\rangle = \sum_{\vk}C_{N\alpha}(\vk-\vk_0;\vx_0)\,|1^{\alpha}_{\vk}\rangle  = \sqrt{N} |\alpha;\vk_0,\vx_0\rangle \,. \label{Nalfa} \ee
Now the localized states (\ref{localsp}) are normalized to the total number of particles in the volume $V$, namely $N=V\sigma^3$, (with one particle per localization volume)   instead of (\ref{normawp}), and they are still orthogonal in the sense of  (\ref{ortomom}).

We will consider the entanglement entropy within a localization volume rather than in the total volume.

  \vspace{2mm}


\subsection{Wigner-Weisskopf with wave packets:}

This wave packet description is easily incorporated into the Wigner-Weisskopf approach to the description of the full time evolution of the quantum state of the decaying parent particle, the quantum state in the interaction picture (\ref{qstate}) is generally written as
\be |\Psi(t)\rangle = \sum_{\vk} C_{\Phi}(\vk,\vk_0;\vec{x}_0;t)|1^{\Phi}_{\vk}\rangle  + \sum_{\kappa} \mathcal{C}_{\kappa}(t) |\kappa\rangle \label{qstate2}\ee where the   states $|\kappa\rangle$ are multiparticle states, with the initial conditions
\be C_{\Phi}(\vk;\vk_0;\vec{x}_0;t=0) = C_{\Phi}(\vk-\vk_0;\vec{x}_0 )~~;~~\mathcal{C}_{\kappa}(t=0) =0 \,, \label{iniwp}\ee where $C_{\Phi}(\vk-\vk_0;\vec{x}_0 )$ describe the localized wave packet of the decaying parent particle at the initial time. The interaction Hamiltonian connects   the single particle plane wave states $|1^{\Phi}_{\vk} \rangle$ with the two-particle plane wave states $|1^{\chi}_{ \vp}\rangle\,|1^{\psi}_{\vk-\vp} \rangle$ with matrix element
\be \mathcal{M}_{\Phi}(\vk,\vp) = \langle 1^\Phi_{\vec{k}} | \hat{H}_I(0)| \chi  _{\vp}\,\psi  _{\vk-\vp}\rangle = \frac{g}{\sqrt{8VE_{\Phi}(k)E_{\psi  }(|\vk-\vp|)E_{\chi  }(p)}} \label{mtxelewp}\ee leading to the decay rate
\be \Gamma_k =   \frac{\Gamma m_\Phi}{E_{\Phi}(k)} \label{gammak} \ee with $\Gamma$ is the decay rate in the rest frame of the parent particle given by eqn. (\ref{gamapi}). Following the same steps as described in the previous section we now find
\be C_{\Phi}(\vk,\vk_0;\vec{x}_0;t) = C_{\Phi}(\vk-\vk_0;\vec{x}_0 )\,e^{-i\Delta E_\Phi(k)\,t}\,e^{-\Gamma_k t/2} \label{cphiwp}\ee where $\Delta E_\Phi(k) = \delta m^2_\Phi/2 E_\Phi(k)$ and
\be \mathcal{C}_{\chi \psi}(\vk;\vp;t) =  C_{\Phi}(\vk-\vk_0;\vec{x}_0)  \,\mathcal{M}_{\Phi}(\vk,\vp)\,\frac{\Big[1-e^{-i(E_{\Phi}(k)-E_\chi  (p)-E_\psi  (|\vk-\vp|)-i\Gamma_k/2)\,t} \Big]}{\Big[E_\psi  (p)+E_\psi  (|\vk-\vp|)-E_{\Phi}(k)+i\Gamma_k/2\Big]}
\,. \label{cchipsiwp}\ee and to leading order in the interaction we find
\be |\Psi(t)\rangle = \sum_{\vk} C_{\Phi}(\vk,\vk_0;\vec{x}_0;t)|1^{\Phi}_{\vk};0_\chi;0_\psi\rangle + \sum_{\vk,\vp}  \mathcal{C}_{\chi \psi}(\vk;\vp;t) |1^{\chi}_{ \vp};1^{\psi}_{\vk-\vp};0^{\Phi} \rangle \,.  \label{qstatet} \ee The number of $\Phi;\psi$ particles respectively are given by
\bea N^{\Phi}(t) & =  & \langle \Psi(t)|\sum_{\vq} a^{\dagger}_{\Phi,\vq}\, a_{\Phi,\vq}|\Psi(t)\rangle = \sum_{\vk}|C_{\Phi}(\vk,\vk_0;\vec{x}_0;t)|^2 \label{Nphi} \\
 N^{\psi}(t) & = & \langle \Psi(t)|\sum_{\vq} a^{\dagger}_{\psi,\vq}\, a_{\psi,\vq}|\Psi(t)\rangle = \sum_{\vp} \sum_{\vk}|C_{\chi,\psi}(\vk,\vp; t)|^2 \,. \label{Npsi}\eea

In order to understand the physical consequences of the wave packet description in the clearest manner, let us consider Gaussian wave packets localized at the origin with vanishing average momentum, namely $\vk_0=0;\vec{x}_0 =0$, and describing a single particle in the volume $V$ so as to establish contact with the plane wave results from the previous section, namely
\be C_{\Phi}(\vk ;\vec{0};\vec{0}) = \Bigg[\frac{8\,\pi^\frac{3}{2}}{\sigma^3\,V} \Bigg]^\frac{1}{2}~e^{-\frac{k^2}{2\sigma^2}}\,.\label{wpzero}\ee

The main assumption in what follows is that the spatial localization length $1/\sigma$ be much larger than the Compton wavelength of the decaying particle $1/m_\Phi$, namely our \emph{main assumption} on the property of the wave packets is that
\be \frac{\sigma}{m_\Phi} \ll 1 \,. \label{assumption}\ee In the case of meson decay this is physically correct as any localization length smaller than the Compton wavelength will necessarily explore the inner structure of the decaying particle  and would be sensitive to the short distance compositeness scale. However as we argue below, this assumption is more general, since a wave packet of massive particles localized on distances shorter than the Compton wavelength will disperse on time scales shorter than the particle's oscillation scale.

Consider the first term in (\ref{qstate2}), it will contribute density matrix elements that will
feature typical integrals of the form
\be I = \Bigg[\frac{8\,\pi^\frac{3}{2}}{\sigma^3\,V} \Bigg]^\frac{1}{2}  \int \frac{d^3 k}{(2\pi)^3} \, e^{-k^2/2\sigma^2} e^{-i\Delta E_\Phi(k)t} e^{-\Gamma_k t/2} \label{integwp} \ee
changing variables to $s = k/\sigma $ it follows that
\be \Delta E_\Phi(k) = \Delta E_\Phi(0)\Big[1- \frac{s^2}{2}\,\frac{\sigma^2}{m^2_\Phi} +\cdots\Big]~~;~~
\Gamma_k = \Gamma \Big[1- \frac{s^2}{2}\,\frac{\sigma^2}{m^2_\Phi} +\cdots\Big]\,, \label{smals}\ee where $\Gamma$ is the decay rate of the particle at rest. The integrand is strongly suppressed for $|s| > 1$  and for $\sigma^2/m^2_\Phi \ll 1 $   the corrections inside the brackets can be systematically computed in a power series expansion, yielding corrections of the form
\be \Gamma t \Big[ \sigma^2/m^2_\Phi + \cdots \Big]~~;~~ -imt\Big[ \sigma^2/m^2_\Phi + \cdots \Big]\,. \label{gcors}\ee The corrections to the decay are negligible during the lifetime of the decaying particle $\Gamma t \simeq 1$   and can be safely neglected.  To lowest order in $\sigma^2/m^2_\Phi$ the corrections to the energy can be absorbed into a time dependent width \be \sigma^2(t) \simeq \frac{\sigma^2} {\big[1+i\frac{\sigma^2}{m^2_\Phi} \big(m_\Phi t)\big]} \label{disp}\ee
 describing the dispersion of the wave packet. For $\sigma^2/m^2_\Phi \ll 1$ we can also neglect the dispersion of the wave packet over the time scale of many oscillations, with the result that
\be I \propto e^{-i\Delta E_{\Phi}(0)t} e^{-\Gamma t/2}\Big[1+\mathcal{O}(\sigma^2/m^2_\Phi)\Big]\,, \label{Iwplo}\ee this is the justification for neglecting the dispersion in the time-evolved wave packet assumed in section (\ref{sec:timeevol}).

Therefore, when considered under integrals (or discrete sums) we can safely replace
\be  C_{\Phi}(\vk;\vec{0};\vec{0};t) \rightarrow   e^{-i\Delta E_{\Phi}(0)t} e^{-\Gamma t/2}\, C_{\Phi}(\vk;\vec{0};\vec{0};0)\,. \label{Cfiwp} \ee

  Physically the result above is the statement that for $\sigma^2/m^2_\Phi \ll 1$ the dispersion can be safely neglected during the lifetime of the decaying state. For example, consider instead that $\sigma/m_{\Phi} \simeq 1$, then the wavepacket will disperse within a time scale given by $m_\Phi t \simeq m^2_\Phi/\sigma^2 \simeq 1$, namely the wavepacket would completely disperse within one oscillation and the concept of the decay rate is not relevant as the amplitude of the wave packet diminishes quickly by dispersion and not by decay.

 In particular to leading order in $\sigma^2/m^2_\Phi$ we find
\be N^{\Phi}(t) = e^{-\Gamma t}~~;~~N^\psi(t) = 1-e^{-\Gamma t}\,, \label{numswp}\ee which are the same results as for the plane wave case. Implementing these approximations, to leading order in this ratio, the first term in  (\ref{qstate2}) becomes
\bea \sum_{\vk} C_{\Phi}(\vk,\vec{0};\vec{0};t)|1^{\Phi}_{\vk}\rangle & = & e^{-i\Delta E_\Phi(0)t}\,e^{-\Gamma t/2}\,\sum_{\vk} C_{\Phi}(\vk,\vec{0};\vec{0};t=0)|1^{\Phi}_{\vk}\rangle + \mathcal{O}(\sigma^2/m^2_{\Phi}) \nonumber \\ & = &   e^{-i\Delta E_\Phi(0)t}\,e^{-\Gamma t/2} |\Phi;\vec{0};\vec{0}\rangle +  \mathcal{O}(\sigma^2/m^2_{\Phi}) \label{wppw} \eea where $|\Phi;\vec{0};\vec{0}\rangle$  is a zero momentum wave packet state localized at the origin.

 Therefore we clearly see that when $\sigma^2/m^2_\Phi \ll 1$, namely when the localization length scale is much larger than the Compton wavelength of the particle we obtain the same result  as in the plane wave case but with the replacement
\be |1^{\Phi}_{\vk} \rangle \rightarrow |\Phi;\vec{k};\vec{x}_0 \rangle \,.\label{replace}\ee

The same argument is applied to integrals of the form
\[ \int d^3 k \,\mathcal{C}_{\chi \psi}(\vk;\vp;t)  \] with $\mathcal{C}_{\chi \psi}(\vk;\vp;t)$ given by (\ref{cchipsiwp}), writing $k = s \sigma$ with the integration range $|s| \lesssim 1$ the k-dependent terms can be expanded around $k=0$, and the k-dependent terms yield corrections in powers of $\sigma/m_{\Phi} \ll 1$, the leading order is given by the $k=0$ contribution, which is obtained by the simple replacement
\be  \mathcal{C}_{\chi \psi}(\vk;\vp;t) \rightarrow C_\Phi(\vk;\vec{0};\vec{0};t=0) \,\mathcal{C}_{\chi \psi}( p;t)\,\label{crepla}\ee where $\mathcal{C}_{\chi \psi}( p;t)$ is the \emph{plane wave result} (\ref{coef}).

The reduced density matrix is obtained by tracing $\rho(t)$ given by (\ref{rho}) over the $\chi$ degrees of freedom, namely
\bea \rho_\psi(t) = \mathrm{Tr}_\chi \rho(t) &  =  & \sum_{\vk,\vk'} C^{\Phi}(\vk;\vec{0},\vec{0},t)\, C^{*\,\Phi}(\vk';\vec{0},\vec{0},t) |1^\Phi_{\vk} \rangle \langle 1^\Phi_{\vk'}| \nonumber \\ & + &  \sum_{\vk,\vk',\vp} \mathcal{C}_{\chi \psi}(\vk;\vp;t)\, \mathcal{C}^{*}_{\chi \psi}(\vk';\vp;t)
|1^\psi_{\vk-\vp}\rangle \langle 1^\psi_{\vk'-\vp}|\,, \label{rhowp}    \eea to leading order in $\sigma^2/m^2_\Phi$ and using the results (\ref{Cfiwp},\ref{crepla}) we find
\bea \rho_\psi(t) & = & e^{-\Gamma t} \,
\Big(\sum_{\vk} C^{\Phi}(\vk;\vec{0},\vec{0},0)|1^\Phi_{\vk} \rangle \Big)\, \Big(\sum_{\vk'} C^{*\,\Phi}(\vk';\vec{0},\vec{0},0) \langle 1^\Phi_{\vk'}| \Big) \nonumber \\ & + &  \sum_{\vp}|\mathcal{C}_{\chi \psi}(p;t)|^2\, \Big(\sum_{\vk } C^{\Phi}(\vk;\vec{0},\vec{0},0)|1^\psi_{\vk-\vp}\rangle\Big)  \Big(\sum_{\vk'}   C^{* \Phi}(\vk;\vec{0},\vec{0},0)
  \langle 1^\psi_{\vk'-\vp}|\Big) \,. \label{rhowpt}    \eea We emphasize that this reduced density matrix was obtained by tracing over the unobserved single particle degrees of freedom $\chi$ \emph{in the plane wave basis}.

  However
  \be \sum_{\vk} C^{\Phi}(\vk;\vec{0},\vec{0},0)|1^\Phi_{\vk} \rangle = |\Phi;\vec{0};\vec{0}\rangle \ee is the original wave packet of the parent particle with zero average momentum and localized at the origin describing a wave packet for the parent particle at rest, and using the property (\ref{prop}) we find
  \be \sum_{\vk} C^{\Phi}(\vk;\vec{0},\vec{0},0)|1^\psi_{\vk-\vp}\rangle = |\psi;-\vp;\vec{0}\rangle \,,\label{dauwp}\ee this is a wave packet of daughter particles with average momentum $-\vp$ localized at the origin with spatial localization length $1/\sigma$ with the same wave packet profile as the parent particle. Namely, the daughter particles ``inherit'' the wave packet structure of the parent particle. Therefore to leading order in $\sigma^2/m^2_\Phi$ we  finally find (after relabelling $-\vp \rightarrow \vp$)
  \be \rho_\psi(t) = e^{-\Gamma t} |\Phi;\vec{0};\vec{0}\rangle \langle \Phi;\vec{0};\vec{0}| +
\sum_{\vp}|\mathcal{C}_{\chi \psi}(p;t)|^2 \, |\psi;\vp;\vec{0}\rangle \langle  \psi;\vp;\vec{0}| \label{psiwpfini}\ee Remarkably, \emph{this is the same result as in the plane wave case} eqn. (\ref{rhonu}) but with the  replacement of the single particle plane wave states by the respective localized wavepackets with the corresponding (average) momenta. The corrections are of $\mathcal{O}(\sigma^2/m^2_\Phi)$ as discussed above.

The function $|\mathcal{C}_{\chi \psi}(p;t)|^2$ is strongly peaked at $p^*$ given by (\ref{pstar}) and has width $\Gamma\ll p^*$. Therefore only wavepackets with momenta $p \simeq p^*$ within a width $\Gamma$ contribute to the reduced density matrix and all of these wavepackets yield the same contribution, just as discussed in the plane wave case. These wave packets are composed of single particle plane wave states with momenta $k\simeq p^*$ within a narrow band of width $\sigma$, for example for the case of pion decay $p^* \approx 30\,\mathrm{MeV}$ and $\sigma \approx 10^{-8}\,\mathrm{eV}$ if this wavepacket is localized within localization lengths $\approx 1\,\mathrm{mt}$. Therefore for a typical experimental value of $p^*$ and macroscopic localization lengths such wavepacket is an excellent approximation to a plane wave with momentum $p^*$.

The density matrix (\ref{psiwpfini}) is in the \emph{interaction picture} where the only time dependence is from the interaction and encoded in the decaying exponential and the  Wigner-Weisskopf coefficients. The density matrix in the \emph{Schroedinger picture} is given by
\be  \rho^{(S)}_\psi(t) = e^{-iH_0t}~ \rho_\psi(t)~ e^{ iH_0t}\,. \label{Sprho}\ee The application of the free time evolution operator on the wave packets yields the wavepackets with the centers displaced by $-\vec{V}_g t$ where the group velocity vanishes for the wavepacket of the decaying particle but it is $\vec{V}_g(p)$ for the $\psi$ wavepackets. Thus the center of the wavepackets moves with the group velocity (again neglecting dispersion).

However, our goal is to obtain the entanglement entropy, for which the unitary transformation (\ref{Sprho}) is irrelevant. Hence insofar as the entanglement entropy is concerned, the motion of the center of the wave packet does not affect the result.



Although the density matrix seems diagonal in the wavepacket description, the wavepackets are not exactly orthogonal for different values of the momenta, so the density matrix in the wave packet representation in principle features off-diagonal matrix elements. However, the coefficient $|\mathcal{C}_{\chi \psi}(p;t)|^2 $ is strongly peaked at a value of the momentum $p^*$ which determined by the kinematics of the decay into plane wave states (the eigenfunctions of the Hamiltonian) from the parent particle at rest, with a Lorentzian profile whose width is the lifetime of the parent particle $\Gamma$. Therefore only states with $p \simeq p^*$ within a region of width $\Gamma$ contribute, typically $p^* ~\mathrm{several}~ \mathrm{MeV} \gg \Gamma$. However, from  the results of section (\ref{sec:local}) above,  wavepackets that are localized on a macroscopic distance  are nearly orthogonal in the sense of eqn. (\ref{ortomom}). Therefore the wave packets furnish an (nearly) orthonormal set and the density matrix is (nearly) diagonal in these states, off diagonal elements only contribute within a width $\sigma$ and their contributions are suppressed by powers of $\sigma/p^* \ll 1$.


The calculation of the entanglement entropy now proceeds just as in the previous section with the final result given by eqn. (\ref{entrofin}).

The fact that the density matrix for the wave-packet description is similar to that of the plane wave description in the regime where the spatial localization scale of the wave packet is much larger than the Compton wavelength of the particle is expected on physical grounds. For example in formal S-matrix theory, the correct approach to describing a scattering event is in terms of localized wave packets for the projectile and target particles prepared in the far past and evolved into the far future. However an actual calculation of a scattering cross section is performed in terms of plane waves, with equal probability everywhere in space. Furthermore, the asymptotic reduction formula that allows to extract S-matrix elements from Green's functions invokes asymptotic in and out states in terms of the single particle eigenstates of the unperturbed Hamiltonian of definite energy and momentum, these being the states that transform as irreducible representations of the Lorentz group. Wave packets do \emph{not} feature definite energy and momenta, yet they underlie all formal descriptions of scattering theory. These two approaches are reconciled when the wave packets of the incoming and outgoing particles are localized on distances larger than the Compton (or de Broglie, whichever is shorter) wavelengths. Similarly, the decay rate of a particle is typically calculated in the plane wave basis, and a particle decaying at rest is assigned a decay rate at zero momentum. However in an experimental situation decaying states are produced generally as wavepackets which are superpositions of plane wave states, each of which would decay with a different rate, yet the description of a decaying particle at rest is in terms of one decay rate extracted as the transition probability per unit time. The decay of a wave packet would entail several different decay time scales. Again this situation is reconciled by considering localized wavepackets but whose localization length is much larger than the typical scale of the particle, namely the Compton or de Broglie wavelength whichever is shorter.

\vspace{2mm}


\subsection{Wave packets of finite particle density:}
In the   treatment of the previous section, we have considered wave packets that describe a single particle in the volume $V$ and explained that this is the reason that the entanglement entropy depends logarithmically on the volume. However, experimentally and more physically we should describe a state that has one particle \emph{ in the localization volume} and not in the total volume. This is achieved by considering the wavepackets describing N-single particles described in section (\ref{sec:Nparts}), namely the states (\ref{Nalfa}). The Wigner-Weisskopf method follows exactly the same steps as before, leading to density matrix in the interaction picture of the same form as (\ref{psiwpfini}) but with the wavepackets $|N\Phi;\vec{0};\vec{0}\rangle= \sqrt{N}| \Phi;\vec{0};\vec{0}\rangle ~;~|N\psi;\vp;\vec{0}\rangle = \sqrt{N}|\psi;\vp;\vec{0}\rangle   $ again with momenta $p\simeq p^*$ within a width $\Gamma \ll p^*$. Now all of the wavepackets with momenta in this interval contribute equally to the sum, with probability $\propto N/V\Gamma \propto  \sigma^3/\Gamma$.   Therefore $\rho(t) \rightarrow N \rho(t)$ leading to the following entanglement entropy
\be S_N(t) = - N \Bigg[ e^{-\Gamma t}\ln\big[N~e^{-\Gamma t}\big]+ \big[1-e^{-\Gamma t}\big]\ln\big[1-e^{-\Gamma t}\big]+ \big[1-e^{-\Gamma t}\big]\ln\big[N n^{\psi  }(p^*\,;\infty)\big] \Bigg] \label{SN}\ee and
\be N n^{\psi  }(p^*\,;\infty) = \frac{4\pi \, m_{\Phi}\,n}{  p^* E_\chi  (p^*) E_\psi  (p^*)\Gamma}\,, \label{Nnpsi} \ee where $n=N/V$ is the particle density. This result is clearly consistent with the discussion on the statistical interpretation of the entanglement entropy in section (\ref{sec:enta}) above, but now with the density $n$.

Assuming that there is one particle \emph{per localization volume} $1/\sigma^3$, it follows that $n=\sigma^3$ and asymptotically the entanglement entropy is given by
\be S_N(\infty) = -N \ln\Bigg[ \frac{4\pi \, m_{\Phi}\,\sigma^3}{  p^* E_\chi  (p^*) E_\psi  (p^*)\Gamma}\Bigg] \,.\label{SNasy}\ee Thus we see, that as anticipated by the discussion above, the entanglement entropy is extensive, the volume dependence has now been replaced by the localization volume, in particular the specific entanglement entropy (entropy per unit total volume) is
\be s_N(\infty) = \frac{S_N(\infty)}{V} = -\sigma^3 \ln\Bigg[ \frac{4\pi \, m_{\Phi}\,\sigma^3}{  p^* E_\chi  (p^*) E_\psi  (p^*)\Gamma}\Bigg]\,,  \label{specSN}\ee and within a localization volume $1/\sigma^3$ we find
\be S_{loc}(\infty) = -  \ln\Bigg[ \frac{4\pi \, m_{\Phi}\,\sigma^3}{  p^* E_\chi  (p^*) E_\psi  (p^*)\Gamma}\Bigg]\,. \label{sloc}\ee

This is the \emph{same} result as in the plane wave case (\ref{entroasy}) but with the localization volume replacing the total volume, which is physically reasonable, for an arbitrary \emph{density} of particles $n=N/V$ the factor $\sigma^3$ is replaced by $n$ and for one particle in the volume $V$ one recovers the result (\ref{entroasy}).


\section{Interpretation and a \emph{possible} experimental measurement}

The logarithmic dependence of the entanglement entropy (\ref{entrofin}) on the volume factor has a clear statistical interpretation. Consider a dilute gas of particles whose statistical distribution  or phase space density is $f_p$. The total \emph{density} of particles is
\be \frac{N}{V} = \int \frac{d^3k}{(2\pi)^3} \, f_p \label{density} \ee and the Von-Neumann entropy of this (dilute) gas is
\be S_{VN} = - \sum_p f_p \ln[f_p]  = - V \int \frac{d^3k}{(2\pi)^3} \, f_p \, \ln[f_p]\,. \label{SVN}\ee  If the number of particles remains finite in the large volume limit, namely if the particle density scales $\propto 1/V$ in this limit, then it follows that $f_p \propto 1/V$. On the contrary,  if $f_p$ is independent of the volume   as in the cases of the Maxwell-Boltzmann, Bose-Einstein or Fermi-Dirac distributions,   the total density is \emph{finite} in the infinite volume limit and the entropy is extensive. For a finite number of particles (vanishing particle density in the infinite volume limit) $f_p \propto 1/V$ and the Von-Neumann entropy is \emph{not} extensive,
\be S_{VN} \propto N \ln[V]\,. \label{nonexSVN}\ee This is \emph{precisely} the origin of the logarithmic dependence on the volume of the entanglement entropy (\ref{entroasy}) for the case of a single particle in the volume $V$: the initial state has one particle and the final state has one (of each) daughter particle, the distribution function of the daughter particles at asymptotically long times after the decay of the parent particle is $|\mathcal{C}_{\chi \psi}(p,\infty)|^2 \propto 1/V$ the inverse volume dependence is the statement that there is a finite number of particles distributed in phase space. Obviously this volume dependence is independent of whether the states are described by  plane waves or wave packets, but is a statement of the simple fact that the number of particles in the volume $V$ is finite.

When we consider wavepackets describing a fixed number of particles \emph{per localization volume}, this case corresponds to a finite density, and for one particle per localization volume it follows that the entanglement entropy becomes extensive and the   volume dependence in the logarithm is replaced by the localization volume $1/\sigma^3$,
\be S_{VN} \propto N \ln[\sigma^3]\,.  \ee
The entanglement entropy from the decay of a parent particle is \emph{in principle}   experimentally accessible: consider the typical experiment in which   a pion is produced at rest from protons incident on a target and the entangled muon-neutrino pairs from pion decay are distributed isotropically. Consider that   muons are detected with a $4\pi$ detector within the pion's decay region but not the neutrinos.   By counting the number of muons within momenta bins of resolution $\Delta p$ the phase space density, namely the muon statistical distribution function is the number of muons detected within this momentum ``bin'' per unit physical detection volume. This is the quantity $f_p$,   the entanglement entropy is the Von-Neumann statistical entropy $S= -\sum_{all ~ bins} f_p \ln[f_p]$ and is a measure of the \emph{information loss} resulting from tracing out the neutrino degrees of freedom and detecting only  muons.
As an order of magnitude estimate consider localization within a detection volume of the order of $(c\tau)^3$ namely with $\sigma = \Gamma$ in eqn. (\ref{sloc}) with $c\tau\approx 7.8\,\textrm{mts}$ being the decay length of the pion at rest, the muon and neutrino are emitted with a Lorentzian distribution of energy peaked at   momenta   $p^* = 30\,\mathrm{MeV}$ and $m_\nu  \ll m_\mu$ and width $\Gamma$ the pion's decay rate, we find the asymptotic entanglement entropy within this volume to be $S_{loc}(\infty) \simeq 70$.

\section{Conclusions and further questions}

The decay of a parent particle leads to a quantum entangled state of the daughter particles as a consequence of conservations laws in the decay process. Experiments at Belle\cite{belle}  measured (EPR) correlations in entangled pairs of $B$ mesons and further experiments at Belle and Babar\cite{trevbo,babar}  exploit the correlations in entangled $B$-meson pairs to study CP and T violation by tagging members of the pairs and studying the time evolution of flavor asymmetries.  Further proposals suggest to extend these studies exploiting entanglement and correlations to measure CP and T violating observables in other $B$-meson systems.

Motivated by these timely experiments that access quantum correlations in entangled states to extract physical information, we focus on a \emph{complementary} aspect of quantum entanglement of particles produced from the decay of a parent particle: if one of the members of the correlated state cannot or will not be measured, tracing over its dynamical degrees of freedom results in a \emph{reduced} density matrix. The Von-Neumann entropy associated with this mixed density matrix, namely the entanglement entropy,  measures the loss of information that was originally contained in the (EPR) correlations of the entangled state.

We generalized and extended a method used in the study of spontaneous decay of atomic systems to the realm of quantum field theory to obtain in a  consistent approximation, the full quantum state that describes the time evolution of the decaying particle and the production of the daughter particles. This method is non-perturbative and is manifestly unitary. We have implemented the method to study the simpler case of bosonic parent and daughter particles to highlight the main concepts and consequences, however, the results are quite general.

The full quantum state resulting from the time evolution of the decaying particle yields a pure state density matrix. However, if one (or more) daughter particles is unobserved, tracing over their degrees of freedom leads to a mixed state density matrix whose time evolution is completely determined by the \emph{unitary} time evolution of the decay process. This mixed state density matrix features an \emph{entanglement entropy} which is a manifestation of the quantum correlations of the entangled product state. We obtained the time evolution of the entanglement entropy and show that it grows on a time scale determined by the lifetime of the decaying particle and reaches a maximum that corresponds to the logarithm of the available phase space states of the decay particles. For a parent particle described by a narrow resonance the distribution of produced (entangled) daughter particles is nearly a constant in a narrow energy-momentum region, the emitted particles are nearly maximally entangled Bell-states.

We have extended the study to the case in which the decaying parent particle is described by a wave packet, rather than a plane wave. The daughter particles ``inherit'' the wave packet structure of parent particle. We have demonstrated the equivalence between the reduced density matrix in terms of plane waves and that in terms of localized wavepackets under the physically reasonable approximation that the localization length $1/\sigma$ of the decaying state is much larger than its Compton wavelength $1/m$ with corrections of $\mathcal{O}(\sigma^2/m^2 )\ll 1$.

Furthermore, we have discussed \emph{possible} experimental ways to access the entanglement entropy. Although unobserved states are manifest as missing energy, the entanglement entropy provides a complementary tool that measures the loss of information contained in the original quantum correlations between the members of the pair of particles produced in the decay process. Just as the measurement of (EPR) correlations at Belle\cite{belle} provide a confirmation of fundamental aspects of quantum mechanics, now accessed at B-factories, a measurement or confirmation of the entanglement entropy could provide yet another complementary test.

While experiments in quantum optics are   testing   fundamental concepts associated with the entanglement entropy and applying them for quantum information platforms\cite{dutt,expt1,photatom}, these concepts have not \emph{yet} received the attention of the particle physics community  but it is conceivable that they \emph{may} prove relevant in the statistical analysis of the time evolution of entangled B-meson pairs in studies of CP violation.

This work is also a  prelude to the assessment of  the entanglement entropy due to particle decays in de Sitter space\cite{desitter}. There, due to the fact that particle can decay into itself with momenta that are much less than the Hubble constant $H_{\rm de Sitter}$\cite{desitter}, we expect there to be an interplay between the horizon size and the decay rate that will feed in to the behavior of the entanglement entropy. This may then mix the ideas of entanglement entropy developed here with those coming from studies of spatially separated portions of de Sitter space such as in ref.{\cite{maldacena}. Work on these aspects will be reported elsewhere\cite{dsnew}.

\acknowledgments D.B. is deeply indebted to  A. Daley,  and D. Jasnow for  enlightening comments and discussions, L. L. and D.B.  acknowledge partial support from NSF-PHY-1202227. R.~H. was supported in part by the Department of Energy under grant DE-FG03-91-ER40682.

\end{document}